\documentclass[9pt,twocolumn,twoside]{opticajnl}
\journal{opticajournal} % use for journal or Optica Open submissions
\setboolean{shortarticle}{true}
\usepackage{graphicx}% Include figure files
\usepackage{dcolumn}% Align table columns on decimal point
\usepackage{bm}% bold math
\usepackage[utf8]{inputenc}
\usepackage[T1]{fontenc}
\usepackage{mathptmx}
\usepackage{etoolbox}

\usepackage{comment}
\usepackage{xcolor}

\usepackage{hyperref}
\hypersetup{
           hidelinks,
           breaklinks=true,   % splits links across lines
           pdfusetitle=true,  % \title and \author values into pdf metadata
        }

\newcommand{\km}{k_\mathrm{m}}

\newcommand{\ko}{k_\mathrm{o}}

\newcommand{\om}{\omega_{\rm m}}

\newcommand{\omopi}{\omega_{\rm m}/(2\pi)}
\newcommand{\oo}{\omega_{\rm o}}

\newcommand{\ooopi}{\omega_{\rm o}/(2\pi)}

\newcommand{\gom}{g_\mathrm{om}}
\newcommand{\nc}{n_\mathrm{c}}
\newcommand{\Com}{C_\mathrm{om}}
\newcommand{\Cem}{C_\mathrm{em}}

\newcommand{\gomopi}{g_\mathrm{om}/(2\pi)}
\newcommand{\gem}{g_\mathrm{em}}
\newcommand{\gemopi}{g_\mathrm{em}/(2\pi)}
\newcommand{\gammaemopi}{\gamma_\mathrm{em}/(2\pi)}
\newcommand{\gammam}{\gamma_\mathrm{m}}
\newcommand{\gammame}{\gamma_\mathrm{m,e}}
\newcommand{\gammami}{\gamma_\mathrm{m,i}}
\newcommand{\gammameopi}{\gamma_\mathrm{m,e}/(2\pi)}

\newcommand{\kappao}{\kappa_{\rm o}}
\newcommand{\kappaoi}{\kappa_{\rm o,i}}
\newcommand{\kappaoe}{\kappa_{\rm o,e}}

\newcommand{\kappamu}{\kappa_{ \mu}}

\newcommand{\etaem}{\eta_\mathrm{em}}

\usepackage{afterpage}
\usepackage{float}

\usepackage{lineno}
\dates{Dated: \today}
\doi{}

\title{Release-free electro-optomechanical crystal modulator}

\author[*]{Paul Burger}
\author[ ]{Joey Frey}
\author[ ]{Johan Kolvik}
\author[ ]{Mads B. Kristensen}
\author[$\dagger$ ]{Raphaël Van Laer}

\affil[ ]{Department of Microtechnology and Nanoscience (MC2), Chalmers University of Technology, 41296 Gothenburg, Sweden}

\affil[*]{paulbu@chalmers.se}
\affil[$\dagger$]{raphael.van.laer@chalmers.se}

\begin{abstract}
Electro-optic modulation is central to classical optical communications and emerging quantum technologies. High-confinement optomechanical crystal modulators enable microwave–optical transduction through strong optomechanical interactions and offer a promising interface between superconducting qubits and optical fibers. However, their performance is limited by thermal noise from optical absorption. Release-free optomechanical crystals provide improved thermal anchoring but have not yet been integrated into a microwave–optical transducer. Here, we demonstrate a release-free electro-optomechanical transducer combining strong optomechanical interactions in silicon with the efficient piezoelectricity of lithium niobate via micro-transfer printing. We observe electro- and optomechanical coupling rates compatible with quantum-level operation when co-integrated with a superconducting microwave circuit. This advance moves release-free electro-optomechanical devices toward practical microwave–optical interfaces.
\end{abstract}

\setboolean{displaycopyright}{false} % Do not include copyright or licensing information in submission.

\begin{document}

\maketitle
\section{Introduction}
Electro-optic modulators connect microwave-frequency electrical signals with optical fields. They combine the low-loss transmission of photons in optical fiber with the strong nonlinearities and high-fidelity processing available in microwave circuits. This interface underpins classical communication systems and is emerging as a key building block for quantum technologies. Today, electro-optic modulators rely on strong microwave drives to establish data links. However, significant efforts aim to push them toward the quantum limit, where individual microwave and optical photons can be interconverted \cite{meesala_nonclassical_2024,jiang_optically_2023, mirhosseini_superconducting_2020,zhao_quantumenabled_2025, sahu_quantumenabled_2022,sekine_microwavetooptical_2025,xie_scalable_2025,brubaker_optomechanical_2022, sahu_entangling_2023}. In this regime, electro-optic interfaces could distribute entanglement between remote superconducting quantum processors, which are leading candidates for error-corrected quantum technologies but face scaling challenges \cite{krastanov_optically_2021, krinner_engineering_2019}. Physical mechanisms enabling the microwave-to-optical photon conversion include the electro-optic effect, the Faraday effect and atomic transitions \cite{han_microwaveoptical_2021}. Electro-optomechanical transducers utilize an intermediary mechanical mode in the transduction chain. They are attractive three-wave-mixing elements given their potential for wavelength-scale co-localization of optical and mechanical fields resulting in strong optomechanical interaction rates, low power consumption and small footprint \cite{chan_laser_2011,safavi-naeini_controlling_2019}. 
In the past years transducers based on suspended optomechanical crystals (OMCs) have become increasingly advanced and are now capable of quantum experiments including optically heralded microwave photon addition \cite{jiang_optically_2023} and non-classical microwave–optical photon pair generation \cite{meesala_nonclassical_2024}. A major challenge is the noise created by heating of the mechanical mode due to parasitic absorption of the optical pump required for the optomechanical three-wave-mixing. This issue is compounded by the fact that state-of-the-art OMCs are suspended, i.e. thermally disconnected from the substrate.
While enabling record mechanical lifetimes \cite{maccabe_nanoacoustic_2020}, the release of the device layer complicates the fabrication and ultimately limits the thermal anchoring. Two-dimensional suspended OMCs \cite{ren_twodimensional_2020, mayor_high_2025,sonar_highefficiency_2025} have shown improved thermal anchoring, though their thermal noise levels remain a limitation and their increased geometric complexity contributes to the challenge of realizing robust transducers.

\begin{figure*}[ht!]
\centering\includegraphics[]{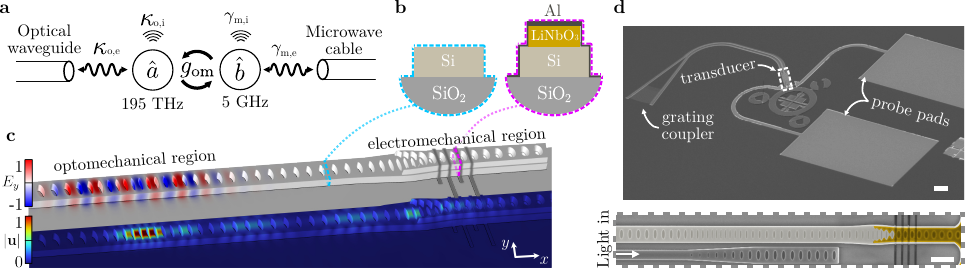}
\caption{\textbf{Release-free electro-optomechanical transducer in silicon-on-insulator.} \textbf{a)} Principle of the transduction process. An optical waveguide is evanescently coupled to an optical mode $\hat{a}$ which optomechanically interacts with mechanical mode $\hat{b}$. The latter is connected to a microwave line through piezoelectric conversion. \textbf{b)} Material stacks in the optomechanical and electromechanical regions. \textbf{c)} Normalized transversal optical field, $E_y$, and mechanical displacement field, $\mathbf{u}$, of the simulated transducer design. \textbf{d)} Scanning electron microscope image of a comparable device (scale bar: $10\ \mu \rm m$). The inset shows a false-colored zoom-in of the electro-optomechanical region (scale bar: $1\ \mu \rm m$).}
\label{fig:overview}
\end{figure*}

Here, we pursue an alternative approach in which the device layer remains attached to a silicon dioxide substrate, improving thermal anchoring and power handling compared to suspended OMCs \cite{kolvik_clamped_2023,kolvik_optomechanical_2025}. With careful choice of the momentum-space operating point it is possible to confine and couple mechanical and optical waves in this release-free architecture, even in presence of the substrate \cite{burger_design_2025}. 
In this work, we experimentally demonstrate a chip-scale release-free electro-optomechanical crystal transducer -- building on the optomechanical component and design developed previously \cite{kolvik_clamped_2023,kolvik_optomechanical_2025,burger_design_2025}. The nanophotonic device is implemented on a silicon-on-$\rm SiO_2$ platform ($\rm SOI$), heterogeneously integrated with thin-film lithium niobate via micro-transfer-printing \cite{carlson_transfer_2012,roelkens_present_2024}. Beyond quantum transduction this device type may be used for applications in classical readout of superconducting microwave circuits and low-power classical electro-optic modulation \cite{vanthiel_optical_2025, miller_attojoule_2017}. As a proof of principle for the transducer as an electro-optic interface, we also demonstrate classical signal transmission.

\section{Design}
In the following, we outline the transducer design briefly. A more detailed description of the design process can be found in \cite{burger_design_2025}. 
The device consists of two regions: In the optomechanical region optical photons and phonons interact via spontaneous parametric up/down-conversion. The process is described by the linearized optomechanical interaction Hamiltonian: $\hat{\mathcal{H}}_{\rm om}=\hbar \gom \sqrt{\nc}( \hat{a}+ \hat{a}^{\dagger}) (\hat{b}+\hat{b}^{\dagger})$, with $\hat{a}$ and $\hat{b}$ the photonic and phononic ladder operators, $\gom$ the vacuum optomechanical interaction rate and $\nc$ the number of photons in the optical mode (Fig. \ref{fig:overview}a) \cite{aspelmeyer_cavity_2014}. 
In the electromechanical region, phonons and microwave photons are converted into one another via the piezoelectric effect at a rate $\gammame$ \cite{jiang_efficient_2020}. The two regions combine to a resonant one-stage transducer, in which optical photons can be converted into microwave photons using phonons as an intermediary \cite{han_microwaveoptical_2021}. To enhance the three-wave-mixing interaction and reduce the power dissipation \cite{safavi-naeini_controlling_2019} the device is realized as an optomechanical crystal cavity.

The presence of the substrate requires revisiting the conditions for mechanical confinement and optomechanical phase-matching \cite{burger_design_2025}. This leads us to use mechanical modes near the X-point, phase protected from the continuum of radiation modes in the substrate. We target optical wavevectors at half the mechanical wavevector, $\km=2\ko$, enabling coupling between a mechanical wave and counter-propagating optical waves in the cavity. The entire device is realized as a beam of silicon which is periodically patterned with elliptical holes to allow tuning the optical and mechanical band structures. The patterning also softens the device layer allowing confinement of the mechanical motion in silicon despite the intrinsically lower speed of sound in the substrate. The electromechanical region additionally features a lithium niobate thin film on top of the silicon along with a short interdigitated capacitor of two periods (Fig. \ref{fig:overview}b-d). While the electromechanical and overall device efficiency could be boosted with a longer electromechanical region, we work with a short region here for compatibility with low-loss superconducting qubits \cite{burger_design_2025,chiappina_design_2023}. We simulate the optical and piezo-mechanical properties using finite-element simulations. The device features a C-band optical mode localized in the optomechanical region and a mechanical mode which is hybridized between both optomechanical and electromechanical regions (Fig. \ref{fig:overview}c). We simulate the vacuum optomechanical coupling rate to be $\gomopi=297 \ \text{kHz}$, while the electromechanical decay rate into a $Z_0=50\ \Omega$ feedline is simulated at $\gammameopi=1.1\ \text{kHz}$.

\section{Measurement}
\label{sec:meas}
\begin{figure*}[ht!]
\centering\includegraphics[]{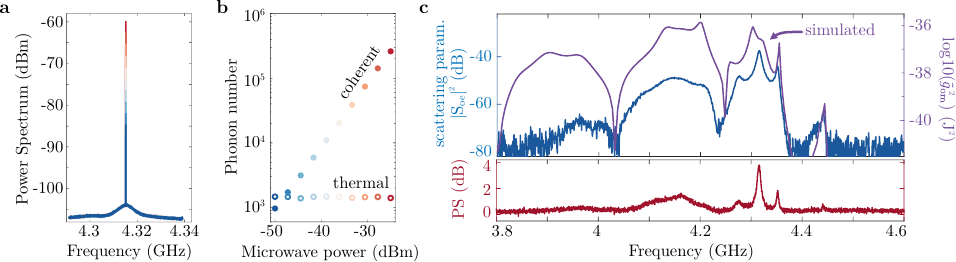}
\caption{\textbf{Room-temperature electro-optic transduction.} \textbf{a)} Mechanical spectrum while a constant frequency microwave tone is applied to the piezoelectric section. The colors correspond to varying microwave power as shown in b). \textbf{b)} Phonon numbers extracted from a) as a function of microwave power. The number of coherently generated phonons rises linearly with power while the thermal population is unaffected. \textbf{c)} Microwave to optical scattering parameter (uncalibrated) as a function of frequency (blue) and simulated response (purple; described in main text and appendix \ref{A3}). The simulated spectrum was shifted by $-160\ \rm MHz$ to align the two. The thermal mechanical spectrum (inset; red; background subtracted) reveals mechanical modes which coincide with peaks in transduction efficiency. Input microwave power $P_{\mu}=-20\ \rm dBm$, not including $\sim 2\ \rm dB$ microwave loss before the device. For panels a)-c) the the pump is blue-detuned ($\Delta=\om$) with $\nc=1.0\cdot 10^{4}$.}
\label{fig:meas2}
\end{figure*}

We fabricate the transducer on a $220\ \rm nm$ silicon-on-$\rm SiO_2$ wafer which we combine with thin-film lithium niobate using micro-transfer-printing (see appendix \ref{A5}). We characterize the $\rm SOI$ release-free transducer at room temperature by routing a tunable laser to the chip through a grating coupler (See appendix \ref{A4} for detailed schematic of the measurement setup). The grating is connected to a bus waveguide which is evanescently coupled to the optomechanical region of the device (Fig. \ref{fig:overview}d). 
We find an optical resonance at $\ooopi=194.9 \ \rm THz$ with optical linewidth $\kappao/(2\pi)=2.1 \ \rm GHz$ of which extrinsic coupling to the bus waveguide accounts for $\kappaoe/(2\pi)=0.99 \ \rm GHz$. The intrinsic optical quality factor is $Q_{\rm o,i}=1.7\cdot10^5$.
Next, we sweep a sideband created with an electro-optic amplitude modulator over the cavity resonance and measure the reflected signal, extracting the detuning between optical pump and cavity frequency, $\Delta=\omega_{\rm L}-\oo$.  With the pump blue-detuned by the mechanical frequency ($\Delta=\om$) we can measure the contributions of thermal phonons modulating the light field with a spectrum analyzer. The spectrum reveals multiple mechanical resonances corresponding to different mechanical modes in the transducer (Fig. \ref{fig:meas2}c, red curve). More exhaustive data of the optical and optomechanical characterization is shown in appendix \ref{A4}. The most prominent mechanical mode is found at $\omopi=4.32 \ \rm GHz$. The device is therefore operated in the resolved-sideband regime, as required for quantum transduction \cite{aspelmeyer_cavity_2014,han_microwaveoptical_2021}. Considering this mode, we measure the mechanical linewidth as a function of intracavity photon number, $\nc$. From the linear fit we extract the vacuum optomechanical coupling rate, $\gomopi=130\ \text{kHz}$, and the intrinsic (backaction-free) mechanical linewidth, $\gammami/(2\pi)=  8.4\ \text{MHz}$. 

We proceed with characterizing the microwave-to-optics conversion. The conversion efficiency \cite{jiang_efficient_2020} of the two-mode system shown in Fig.~\ref{fig:overview}a is given by:
$$\eta_{\rm tot}=\eta_{\rm o,c}\eta_{\rm o}\etaem\frac{4\Com}{(1\mp \Com)^2} \quad \text{with} \quad \Com=\frac{4\nc\gom^2}{\kappao\gammam}.$$ 
Here, $\eta_{\rm o, c}=0.29$ and $\eta_{\rm o}=\kappaoe/\kappao=0.47$ are the efficiencies of the grating and the bus-to-cavity coupling respectively. $\etaem=\gammame/\gammam$ is the efficiency of the coupling of the hybridized mechanical mode to the microwave feedline \cite{jiang_efficient_2020,jiang_lithium_2019}. The $-(+)$ sign in the denominator corresponds to blue (red) detuned operation.\newline
We bring a microwave probe attached to a $50\ \Omega$ transmission line into contact with the aluminum pads on the chip making sure to punch through any surface oxide. Applying a microwave tone at a fixed frequency coherently drives the piezoelectric transducer. The phonons created by the piezoelectric interaction contribute to the optomechanical sideband which we measure on the optical side in the regime of low cooperativity, $\Com \ll 1$. 
The mechanical spectrum measured as before now shows a sharp peak on top of a wide lorentzian associated with the coherent and the thermal phonons respectively (Fig.\ref{fig:meas2}a, b). 
Using the thermal mechanical peak as a calibration value we can determine the number of coherent phonons which in turn allows us to deduce the mechanical to microwave decay rate $\gammaemopi=58\ \text{Hz}$ \cite{jiang_lithium_2019}. This implies an electromechanical conversion efficiency of $\etaem=7\cdot 10^{-6}$. 
We also conduct a sweep in microwave power and record the spectra and phonon numbers (Fig. \ref{fig:meas2}b). 
The frequency dependency of the transduction from microwave to optical domain, $S_{\rm oe}$, is measured with a vector network analyzer (Fig.~\ref{fig:meas2}c, blue line). The response shows a band at around $4.3\ \rm GHz$ with multiple transduction peaks. Comparison to the thermal mechanical spectrum (red line) reveals that each transduction peak corresponds to a mechanical mode in the device which is also optomechanically responsive. 
Based on feedback from measurement and SEM we simulate the fabricated device geometry---which differs from the initial design---and recreate the spectrum (purple line). The qualitative agreement allows us to identify the mechanical modes contributing to the transduction (see appendix \ref{A3}).
The spectrum shown in Figure \ref{fig:meas2}c is taken at on-chip optical power of $P_{\rm opt}=-7.9\ \text{dBm}$, corresponding to $\nc=1.0\cdot 10^{4}$. This implies a total microwave-to-optical conversion efficiency of $\eta_{\rm tot}=1.5\cdot 10^{-7}$. We note that, since the pump is blue-detuned, the process used here does not correspond to direct microwave-to-optical photon conversion — which would be realized with a red-detuned pump. Rather, the piezoelectrically generated phonons stimulate parametric pair generation, producing correlated optical sideband photons.

%\begin{comment}
\section{Classical signal transmission}
\label{sec:bit_array}
%Next, we test the transducer as the electro-optic interface in a classical communication datalink. 
To complement the microwave-to-optical characterization in Sec. \ref{sec:meas}, we now demonstrate classical data transmission through the transducer as a proof-of-principle of its functionality as an electro-optic interface.
Our transducer is resonant in both the optical and acoustic domains. This means that the digital radiofrequency signal at the input needs to be first upconverted to the transduction band at the mechanical frequency $f_{\rm m}$ (Fig. \ref{fig:bitstring}a; see appendix \ref{A6} for more information).

\begin{figure}[ht!]
\centering\includegraphics[]{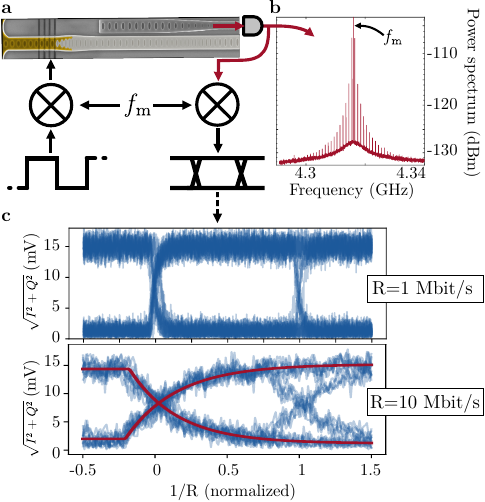}
\caption{\textbf{Classical electro-optomechanical transmission of bit array.} \textbf{a)} Simplified schematic of the measurement scheme used in place of the VNA employed for the microwave-to-optical characterization in Sec. \ref{sec:meas} (appendix \ref{A4}). For acousto-optic modulation an arbitrary digital signal with amplitude $V_0=100\ \text{mV}$ is up- and downconverted to/from the transduction band. \textbf{b)} The power spectrum of an upconverted square pulse train with fundamental frequency $0.5\ \rm MHz$ on top of the thermal optomechanical signal. \textbf{c)} Eye diagrams of a non-return-to-zero bit array measured in time domain after demodulation for two different bit rates $R$. The red lines show fits to exponential ring-up and ring down.}
\label{fig:bitstring}
\end{figure}

When the microwave signal is applied to the electromechanical side of the transducer the arising phonons modulate the optical carrier. With the device read out on the optical side as before we can measure the microwave power spectrum of the modulated optical output. With a square waveform at the radiofrequency input, for instance, the spectrum shows the characteristic peaks at the odd harmonics around the local oscillator frequency $f_{\rm m}$ (Fig. \ref{fig:bitstring}b). Analogously, any arbitrary bit array can be encoded using the transducer. For a non-return-to-zero array with 48 bits we record the transduced signal after downconversion in time domain. Aligning the rising and falling edges, corresponding to switches between logical 1 and 0, produces a characteristic eye diagram (Fig. \ref{fig:bitstring}c).  With increasing bit rate $R$, frequency components of the signal begin to exceed the transduction bandwidth which decreases the extinction. For $R\sim \gammam$ the mechanical ring-up and ring-down becomes prominent which ultimately limits the bandwidth of the modulation. For $R=10\ \text{Mbit/s}$ we fit the datasets to an exponential function, $\sim\exp(-\gammam  t/2)$, from which we receive $\gammam/(2\pi)=9.25\ \text{MHz}$, deviating around $15\%$ from our estimate of the total mechanical linewidth $\gamma_{\rm tot}/(2\pi)=7.9\ \text{MHz}$ (see appendix \ref{A6}). 
%\end{comment}
\section{Discussion and outlook}
The measured  $\gom$ and especially $\gammame$ deviate from those predicted by simulations. We attribute this to discrepancies between the intended design geometry and the dimensions extracted from scanning electron microscope (SEM) images. We use the SEM images to refine our simulations after the measurements finding qualitative agreement with the observed spectrum (see appendix \ref{A3}).
Despite this circumstance we find appreciable performance owed to the robustness of the design \cite{burger_design_2025}. 
 
The measured electromechanical conversion efficiency \mbox{$\etaem=7\cdot 10^{-6}$} is set by the intrinsic mechanical loss of the device,  $\gammami$ , and the electromechanical coupling,  $\gammame$. While lowering $\gammami$ may not be beneficial by default as it also influences the transducer's bandwidth and added noise, increasing the coupling $\gammame$ is certainly desirable. Depending on the application at hand, this could be addressed by increasing the size of the electromechanical region while adding more periods to the interdigitated capacitor. The resulting increase in the IDT capacitance , $C_{\rm IDT}$, enhances the microwave participation in the electromechanical region and, together with increased mechanical participation, leads to a linear improvement in the electromechanical interaction strength, $\gammame\propto C_{\rm IDT}$.

For operation in the quantum regime, a dramatic boost of microwave-to-mechanical conversion can be realized by coupling the transducer to a microwave resonator or a superconducting qubit. In this case, the efficiency $\etaem$ increases with the impedance of the microwave resonator as well as the quality factors of the microwave and acoustic modes, which benefit from cryogenic temperatures \cite{jiang_lithium_2019}. For a transmon qubit of typical performance the electromechanical cooperativity then becomes high enough, $\Cem\gg1$, such that the transduction efficiency is no longer limited by the electromechanical performance (see appendix \ref{A7} and \cite{burger_design_2025}). In particular, with the electromechanical coupling rate measured (simulated) here, our transducer supports Rabi swapping operations to a superconducting qubit if the cryogenic mechanical loss rate drops below $\gammami/(2\pi) < 3\ \rm MHz$ ($<14\ \rm MHz$). Instead, the optomechanical section becomes the bottleneck as parasitic heating of the mechanical mode strongly limits the available optical power. We expect the release-free approach with its improved heat sinking to help relax this trade-off between noise and optomechanical cooperativity or conversion efficiency \cite{burger_design_2025,kolvik_optomechanical_2025}. 
Separately, integrating superconducting circuits with the transducer is challenged by quasiparticle poisoning induced by stray light \cite{mirhosseini_superconducting_2020}. Possible mitigation strategies include developing light-resilient qubits or housing them in a separate light-shielded package.

High-confinement acousto-optic devices are also an interesting platform for energy efficient classical electro-optic modulation \cite{safavi-naeini_controlling_2019,miller_attojoule_2017}. In Sec. \ref{sec:bit_array} we demonstrated a simple data transmission experiment without making use of techniques like filtering or data post-processing which would improve the data quality at the receiving end. In principle, silicon-based optomechanical interactions could unlock attojoule-per-bit power consumption with resonant microwave and optical enhancement, thanks to the wavelength-scale overlap of the optical and mechanical fields \cite{safavi-naeini_controlling_2019,jiang_efficient_2020}. Our release-free platform could also excel in applications where efficient heat sinking \cite{kolvik_optomechanical_2025} is required, such as in classical and quantum readout of superconducting microwave circuits \cite{vanthiel_optical_2025, mirhosseini_superconducting_2020}. 

In summary, we have presented the fabrication and characterization of the, to the best of our knowledge, first release-free piezo-optomechanical transducer based on a high-confinement optomechanical crystal. 
Refining the fabrication process should allow us to realize dimensions closer to the simulated design and improve figures of merit such as the optomechanical interaction rate $\gom$ and the electromechanical conversion efficiency $\etaem$. Together with the use of appealing substrates such as sapphire \cite{burger_design_2025} and novel design techniques such as inverse design \cite{hambraeus_inversedesigned_2026}, this platform provides a new route toward the co-integration of phononic, photonic, and superconducting systems. Operating the device under cryogenic conditions would enable us to evaluate the performance at the single-phonon level bringing quantum experiments using superconducting qubits into reach. 

\begin{backmatter}

\bmsection{Acknowledgment} We thank Trond Haug for helpful discussions and Niclas Lindvall for assistance with the aligned lithography. This work was performed in part at Myfab Chalmers. 

\end{backmatter}

%\section{References}

% Bibliography
\bibliography{PB_ZoteroLib_2026}

% Full bibliography added automatically for Optics Letters submissions; the following line will simply be ignored if submitting to other journals.
% Note that this extra page will not count against page length
%\bibliographyfullrefs{PB_ZoteroLib_2026}
\newpage
\section*{Appendix}
\smallskip
\appendix

\section{Device parameter overview}
\label{A1}
\begin{table}[ht!]
    \caption{\textbf{Key parameters of the presented release free SOI transducer.} The rows show measured values along with the corresponding simulated ones. \textit{Sim. (init.)} denotes the values predicted by the initial simulation underlying the optimized design. \textit{Sim. (adj.)} denotes those calculated based on the simulation adjusted for deviations observed in the fabrication (see sec. \ref{sec:si_sim}). The simulated intrinsic loss rates, indicated by parenthesis, are estimated based on radiative loss and do not take other effects such as fabrication disorder into account. The intrinsic mechanical loss of the adjusted simulation, \textit{Sim. (adj.)}, does additionally include intrinsic material loss as described in Sec. \ref{sec:si_sim}.}
    \setlength{\tabcolsep}{10pt} % Default value: 6pt
    \centering
    %\begin{ruledtabular}
    \begin{tabular}{l |c| c|c }
           & Meas.  & Sim. (adj.) & Sim. (init.) \\ \hline
        $\oo/2\pi\ $ (THz)  & 194.9  & 193.7 & 195.1 \\
        $\kappao/2\pi\ $ (GHz) & 2.1  & - & - \\
        $\kappaoe/2\pi\ $ (GHz)  & 0.99& -  & - \\
        $\kappaoi/2\pi\ $ (GHz)  & 1.12  & (0.55) & (0.13)\\\hline
        $\om/2\pi\ $ (GHz) & 4.32  & 4.46 & 4.06\\ 
        $\gammami/2\pi\ $ (MHz)  & 8.4  & (9.4) & (3.2)\\\hline
        $\gammame/2\pi\ $ (Hz)  & 58 & 186 & 1140\\
        $\gom/2\pi\ $ (kHz)  & 130  & 139 & 297\\
        $C_{\rm IDT}$ (fF) & - & 0.42 & 0.41\\
    \end{tabular}
    %\end{ruledtabular}
\label{tab:meas_summary}
\end{table}

\section{Design geometry parameters}
\label{A2}
\label{sec:designParams}
The design process of the release-free transducer presented in this work is extensively described in \cite{burger_design_2025}. In contrast to the silicon-on-sapphire device in \cite{burger_design_2025}, the substrate layer of the measured device here, silicon-dioxide, has a \textit{lower} speed of sound than the device layer. Therefore, the design relies on geometric softening to lower the frequency of the mode of interest in the silicon below the acoustic modes in the substrate \cite{kolvik_clamped_2023}.

In the following tables, Tabs. \ref{tab:paramOMC}, \ref{tab:paramEMC}, \ref{tab:paramN}, we provide the relevant parameters for the different sections of the device. Their exact role in the design can be found in \cite{burger_design_2025}. These numbers refer to the initial design, which are then subject to deviations during the fabrication process (see Sec. \ref{sec:si_sim}).

\begin{table}[h!]
\caption{\textbf{Parameters of the OMC region.}  The silicon device layer is 220 nm thick.}
    \centering
\begin{tabular}{ c| c| c| c } 
    & Mirror (nm)& Defect (nm)& Partial mirror (nm)\\ \hline
   $a$  & $386 $  & $ 198 $  & $ 348$  \\
    $h_{x}$  & $202 $   & $97 $  & $145 $ \\
    $h_{y}$  & $ 460$   & $ 448$  & $393 $ \\
    $w$  & $700$   & $700$  & $700$ \\
\end{tabular}
\label{tab:paramOMC}
\end{table}

\begin{table}[h!]
\caption{\textbf{Parameters of the EMC region.} The lithium niobate device layer is $150\ \rm nm$ thick. The tabulated values $h_x$, $h_y$, $w$ are given at the top of the lithium niobate layer. The ellipse values at the top of the silicon layer can be retrieved by calculating the sidewall width and allowing for a 30 nm  (50 nm) misalignment buffer in $h_x$ ($h_y$) direction of the hole. The sidewall angles are 18° and 10° in the hole and at the beam edge respectively. The silicon beam is assumed to have 0° sidewall angle. The taper-end cell (Si) column gives the values at the top of the silicon layer. The polynomial $f$ describing how $h_x$ is transformed in the transition region is: $h_x=f(h_y)=\exp(a\cdot h_y+b)-\exp(-(a\cdot h_y+c))+d$. The parameters are $[a,b,c,d]=[ 0.0344 \ \text{nm}^{-1},\ 	-17.84\ ,\	-17.27\ ,\	256.86\ \text{nm}]$.}
    \centering
\begin{tabular}{ c | c |c |c |c } 
 & Taper-End (Si) (nm) & Taper-End (nm)&Defect (nm)& Mirror (nm)\\  \hline
   $a$  & $190 $  & $ 190$  & $ 290$  & $ 375$  \\
    $h_{x}$   & $95 $  & $f$  & $188 $  & $249 $  \\
    $h_{y}$  & $462 $  & $ 622 $   & $ 377$  & $ 372$    \\
    $w$  & $ 840$  &  $ 710$   & $ 570$  & $570 $   \\
\end{tabular}   % starting rightmost sub table
\label{tab:paramEMC}
\end{table}

\begin{table}[h!]
\caption{\textbf{Number of unit cells in each region} The table lists the number of transition cells between adjacent regions.}
    \centering
\begin{tabular}{ c|c } 
%\hline
   Device region  & $N_{\rm uc}$  \\  \hline
    OMC mirror - OMC defect  & $10$  \\
   OMC defect - OMC partial mirror  & $10$  \\
   OMC partial mirror  & $7$  \\
   OMC partial mirror - taper end cell  & $5$  \\
   taper end cell - EMC defect & $5$   \\
   EMC defect - EMC mirror  & $5$  \\
\end{tabular} \\ 
\label{tab:paramN}
\end{table}

To address the transducer's EMC we use 4 aluminum electrodes, i.e. 2 periods, 80 nm in width and 20 nm in thickness.

As described in the main text and sec. \ref{sec:si_sim}, the fabricated and measured device differs in geometry compared to the initial design. In sec. \ref{sec:si_sim} we state the geometrical changes used to recreate the measured spectrum. 

\section{Simulated microwave-Optical transduction spectrum $S_{\rm oe}$}
\label{A3}
\label{sec:si_sim}
The microwave-to-optical transduction spectrum exhibits numerous conversion bands and peaks above the noise floor (main text Fig. 3c). To gain understanding into the underlying excited mode profiles we conduct frequency domain simulations of the full transducer. In the simulation a fixed microwave power $1\ \mu \rm W$ from a $50 \, \Omega$ microwave port is applied to the electrodes. We then calculate the optomechanical overlap integral between the resulting mechanical mode and the optical mode at varying drive frequencies without considering normalizations. In other words, we calculate $\tilde{g}_{\text{om}}$, which for the (dominating) moving boundary contribution is:
\begin{equation}
\label{eqn:g_rp}
\begin{aligned}
      \gom|_{\mathrm{m.b.}}&=\sqrt{\frac{\hbar}{2 \omega_{\mathrm{m}}}} \frac{\omega_{\mathrm{o}}}{2}\frac{\tilde{g}_{\text{om}}|_{\mathrm{m.b.}}}{\sqrt{\int \rho|\boldsymbol{u}(\boldsymbol{r})|^2 \mathrm{~d}^3 \boldsymbol{r}} \int \epsilon(\boldsymbol{r})|\boldsymbol{E}(\boldsymbol{r})|^2 \mathrm{~d}^3 \boldsymbol{r}},\quad\text{where} \\ \tilde{g}_{\text{om}}|_{\mathrm{m.b.}}&=\int(\boldsymbol{u}(\boldsymbol{r}) \cdot \boldsymbol{n})(\Delta \epsilon|\boldsymbol{E}^{\|}|^2-\Delta(\epsilon^{-1})|\boldsymbol{D}^{\perp}|^2) \mathrm{d} A.
 \end{aligned}
\end{equation}
The result of the integral, $\tilde{g}_{\text{om}}$, is meant to reflect a transduction efficiency relative to that at other driving frequencies. 

There are significant differences between the intended design and the fabricated device. These differences include the size of the elliptical holes, the sidewalls of the silicon and the electrode thickness. Using feedback from scanning electron microscopy and simulations we aim to recreate the measured spectrum by adjusting the simulated geometry. We also include intrinsic mechanicalloss by introducing an isotropic loss factor to the materials. Each material's loss factor is chosen such that a mechanical mode fully confined to it would be limited to a certain intrinsic quality factor. For the implementation available in \textit{COMSOL multiphysics} this corresponds to $\eta=1/Q_{\rm m,i}$. For our room temperature measurements we choose those to be $Q_{\rm m,i}=(2500,\ 2500,\ 300)$ for silicon, lithium niobate and aluminum respectively. The mechanical modes in the transducer then experience loss depending on their energy participation across those domains.

\begin{figure}[ht!]
\centering\includegraphics[]{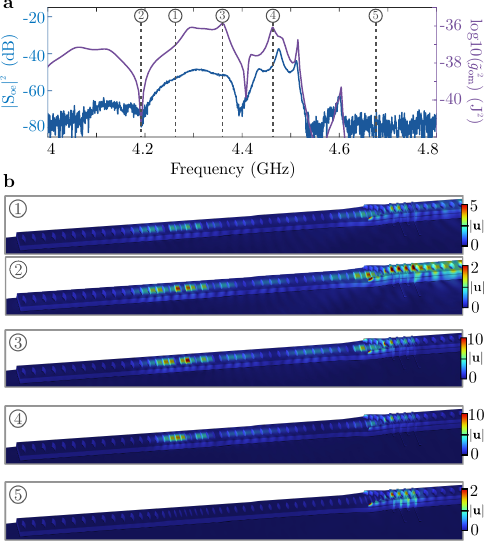}
\caption{\textbf{Microwave-to-optics transduction.} \textbf{a)} Measured (blue) and simulated (purple) microwave-to-optical scattering parameter. The measured data was shifted by $+160\ \rm MHz$ to align the spectra. The measured $S_{\rm oe}$ is uncalibrated. The simulated $S_{\rm oe}$ and its origin is described in the text. \textbf{b)} Mechanical mode profiles corresponding to selected driving frequencies in a). Colorbar unit: $\text{pm}$.}
\label{fig:sim_Soe}
\end{figure}

Figure \ref{fig:sim_Soe} shows the measured and simulated scattering parameters along with the mode profiles. The measured data was shifted by $+160\ \rm MHz$ to align the general characteristics of the two spectra, such as peaks, dips and cutoff frequency. We find reasonable qualitative agreement indicating that even the transduction bands measured away from the main peaks are of piezo-optomechanical origin, and not e.g. electro-optic. The sharp cut-off at $4.5\ \rm GHz$ coincides with an emerging quasi-bandgap at that frequency in the partial mirror region of that device. At drive frequencies around $4.2\ \text{GHz}$ and $4.4\ \text{GHz}$ the transmission drops significantly. Mechanical mode profiles excited at those frequencies are shifted such that they are aligned unfavorably to the optical mode profile which causes optomechanical coupling of cells to add up destructively. \newline
This adjusted geometry yields different simulation results compared to the initial simulation presented in the design section of the main text (see also tab. \ref{tab:meas_summary}). For example, there is now a higher order optical mode near 1550 nm compared to the initial design. In the eigenmode simulation of the adjusted design the optical frequency is $\ooopi=193.7\ \rm THz$ and the optomechanically strongest mechanical mode is at $\omopi=4.46\ \rm GHz$. The corresponding coupling rates are $\gomopi\approx 139 \ \text{kHz}$ and $\gammameopi= 186\ \rm Hz$. We therefore suspect that the differences between fabrication and design contributed to the lower-than-expected coupling rates in measurement.
The adjusted simulation only captures a subset of factors contributing to the discrepancy between measurement and design. For example, we still observe residual detuning, $\approx20\ \text{MHz}$, between the two most prominent modes in Fig. \ref{fig:sim_Soe}, and the simulated $\gammame$ exceeds the measurement significantly. Notably, the SEM images informing the simulations were taken of a comparable device and not of the measured device in order to avoid damage caused by the SEM. Other contributing factors could include an oxide layer in between the LN and silicon layers or misalignment between the transfer printed electromechanical region and the IDTs or the rest of the device.\newline

In order to match the final fabricated device in simulation, we make make the following adjustments to the initial design parameters shown in sec. \ref{sec:designParams}:
\begin{itemize}
    \item The measured device was part of a parameter sweep on the chip: The EMC defect cell has a factor of 1.014 applied to $a$ and $h_y$. The partial mirror cell has a factor of 0.98 applied to $h_y$.
    \item The sidewall angle of the silicon layer is assumed to be 7° at the outer edge. Inside the holes the angle is a logistic function of the hole size: 
    \begin{equation}
        \alpha(h_x)=\alpha_1+ \frac{\alpha_2-\alpha_1}{1+\exp \big(-\frac{h_x-(x_0+x_1)/2}{(x_1-x_0)/2}\big)}, 
    \end{equation}
     where $\alpha_1=1.5°$, $\alpha_2=4.5°$, $x_0=90\ \text{nm}$ and $x_1=150\ \text{nm}$.
     \item To account for the size differences of the elliptical holes in the silicon we apply a function to the ellipse parameters,\newline $f(x)=a+b\cdot\ln(1+\exp(c\cdot(x-d))$. 
     \begin{itemize}
         \item $h_x$: $[a,b,c,d]=[45\ \text{nm},\ 8.84\ \text{nm},\ 0.12\ \text{nm}^{-1},\ 77.5\ \text{nm}]$
         \item $h_y$: $[a,b,c,d]=[45\ \text{nm},\ 6.1\ \text{nm},\ 0.146\ \text{nm}^{-1},\ 21.8\ \text{nm}]$
         \item $w$: $f(x)=1.029\cdot x$.
     \end{itemize}
     \item The electrode thickness is 50 nm.
     \item The silicon layer below the EMC is not through etched.
\end{itemize}

\section{Optical and Mechanical characterization}
\label{A4}
\begin{figure*}[ht!]
\centering\includegraphics[]{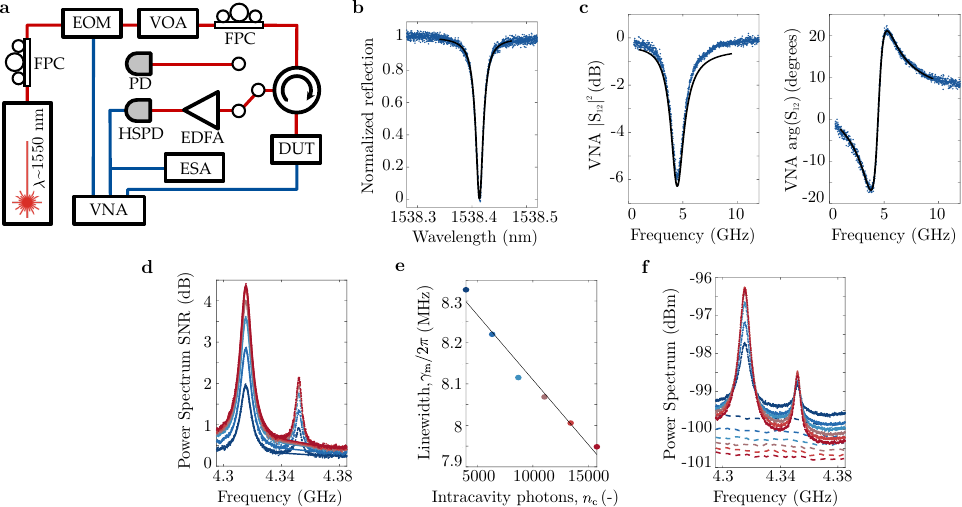}
\caption{\textbf{Room-temperature optical and optomechanical characterization} \textbf{a)} Overview of the measurement setup. FPC: fiber polarization controller. EOM: electro-optic amplitude modulator. VOA: variable optic attentuator. EDFA: erbium-doped fiber amplifier. PD: photo-detector. HSPD: high-speed photo-detector. ESA: electronic spectrum analyzer. VNA: vector network analyzer. DUT: device under test. \textbf{b)} Normalized optical cavity reflection of laser swept over the resonance. \textbf{c)} Magnitude and phase response of the reflected light when sweeping a sideband over the cavity resonance. After subtracting the background measured far away from the resonance, we fit the phase response (black line) to extract the detuning $\Delta$. \textbf{d)} Microwave power spectrum of the reflected light showing the mechanical modes for a blue detuned pump, $\Delta=\om$. The colors correspond to varying intracavity photon number as shown in d). The spectrum is plotted in units of signal-to-noise ratio with respect to the background measured with the laser far detuned. \textbf{e)} Linewidths extracted from c) as a function of intracavity photon number. \textbf{f)} Raw data of the mechanical power spectrum shown in the main text. The dashed line indicates the background measured when detuning the laser far away from the cavity, $\Delta\approx10 \kappao$.}
\label{fig:measSI}
\end{figure*}

Figure \ref{fig:measSI}a shows a schematic of the measurement setup used for the measurements in the main text. Also presented is the optical cavity transmission spectrum (Fig. \ref{fig:measSI}b), received by sweeping the tunable laser over the cavity resonance. 
For measurements involving the laser detuned from the cavity, we sweep a sideband created with an electro-optic amplitude modulator over the cavity resonance and measure the reflected signal with a high-speed photodetector. The interference between the sideband and the pump gives rise to GHz components which we read out with a vector network analyzer. By fitting the phase response we determine the detuning, $\Delta=\omega_{\rm L}-\oo$ (Fig. \ref{fig:measSI}c)\cite{chan_laser_2012}.
Figs. \ref{fig:measSI}d-f show the optomechanical characterization, where we sweep the optical power while recording the mechanical linewidth. This measurement is used to determine $\gom$, as described in the main text.

\section{Device fabrication}
\label{A5}
\label{app:fab}
\begin{figure*}[ht!]
\centering\includegraphics[]{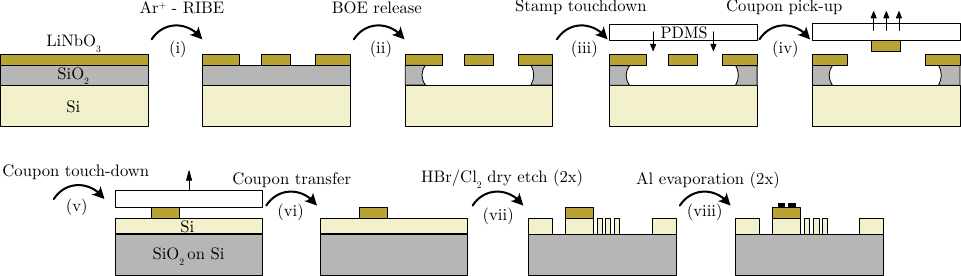}
\caption{\textbf{Full transducer fabrication flow.} (i) A commercial LN-on-Insulator (LNOI) chip is patterned, defining the electromechanical region and the micro-transfer printing coupon. (ii-iv) The LN device layer is suspended and then picked up with a PDMS stamp. (v-vi) The LN device is transfer printed onto the $\rm SOI$ target chip. (vii) After annealing we perform two successive lithography and dry etch steps to define first alignment marks and then the silicon pattern. (viii) Finally, aluminum electrodes and microwave probe pads are deposited on the device.}
\label{fig:fab}
\end{figure*}

We integrate X-cut lithium niobate thin-film devices (LN) with the $\rm SOI$ substrate using micro-transfer-printing (Fig. \ref{fig:fab}).
Starting from a commercial LN-on-insulator chip, we use argon ion beam etching (IBE) to mill down the LN to $150\ \rm nm$, which we confirm with ellipsometry. We then define the electromechanical device region using electron beam lithography (EBL) and Hydrogen silsesquioxane (HSQ) resist. The pattern includes a large cross which will allow us to align the following lithography layers to the transfer-printed device (main text Fig. 1d). After etching with IBE the chip is cleaned with buffered oxide etch (BOE), piranha solution and heated hydrofluoric acid (HF). We suspend the LN layer by wet-etching the sacrificial oxide layer underneath with BOE. Using a Polydimethylsiloxane (PDMS) film mounted to a set of manual translation stages the device is then picked up and transferred to a $220\ \rm nm$ $\rm SOI$ chip (Frey et al., manuscript in preparation). To promote the LN-Si bond the chip is then annealed overnight at $500\ \rm ^{\circ} C$ and cleaned of PDMS residues with piranha solution.\newline
To account for displacement and rotation of the transferred devices subsequent lithography layers are aligned to the device. The first post-transfer layer defines fresh alignment crosses in the silicon but deliberately does not include the optomechanical region to avoid exposure to inadvertently etched lithium niobate. We proceed with the next layer creating the sensitive silicon parts of the device using multipass EBL and plasma etching with $\rm HBr/Cl_2$ chemistry. The sample is cleaned with a single cycle of $3:1$ piranha solution and a dip in 2\% hydrofluoric acid. 
While multiple piranha/HF cycles could improve surface roughness and thus optical and mechanical performance, it also poses a risk to the LN adhesion and gradually etches the SiO$_2$ substrate which is not desired.
Finally, the interdigitated transducer (IDT) and the probe pads are defined with angled and top-down aluminum evaporation followed by lift-off. This is done in two separate layers because of the large size difference between the structures, the different desired thicknesses for IDT (50 nm) and pads (170nm), as well as potential challenges when lifting off the pads after an angled evaporation. Just before the second evaporation we argon-mill around 20 nm into the first layer to remove any insulating native oxide. The finished device is shown in main text Fig. 1d.

\section{Bit array measuremement}
\label{A6}
To conduct the experiment we use a Quantum Machines OPX and Octave system which act in concert to provide waveform generation as well as up- and down-conversion. These replace the VNA in the setup shown in Fig. \ref{fig:measSI}a. We first generate a signal at an intermediate frequency, $f_{\rm IF}=50\ \text{MHz}$, to which a square shaped waveform corresponding to the given bit array is applied. This signal is then upconverted to the mechanical frequency using an LO at $f_{\rm LO}=f_{\rm m}-f_{\rm IF}$. This mixing process is calibrated to suppress unwanted sidebands. The output signal is thus centered around $f_{\rm m}$ (main text Fig. 3b). The voltage amplitude is $V_0=100\ \text{mV}$ corresponding to an average power $P_{\rm avg}=\frac{1}{4}\frac{V_0^2}{50\Omega}=-13\ \text{dBm}$. We estimate around $2\ \text{dB}$ of microwave loss to the device. 

We detect the optical signal using the same setup as for the microwave-to-optical characterization. The optical pump is blue detuned ($\Delta=\om$) with an on-chip power of around $P_{\rm opt}=-5\ \text{dBm}$, corresponding to $\nc=1.8\cdot 10^{4}$. In this regime of low optomechanical cooperativity, $\Com=0.07$, the impact of the optomechanical interaction on the total mechanical linewidth is small, $\gamma_{\rm tot}=\gammami-\gamma_{\rm om}$, with $\gammami/(2\pi)=8.4\ \text{MHz}$ and $\gamma_{\rm om}=4\nc\gom^2/\kappao= 2\pi\cdot 540\ \rm kHz$.

After detection of the optical signal, the microwave signal is amplified to lift it into the detection range of the Octave. Once demodulated with the LO, the complex voltage signal $V_\mathrm{det}(t)$ is given by the quadratures $V_\mathrm{det}(t) = I(t) + iQ(t)$ where $I(t)\propto\cos(2\pi f_{\rm IF}t)$ and $Q(t)\propto\sin(2\pi f_{\rm IF}t)$. We note, that the voltage given in main text Fig. 4c is not measured right after the transduction but after a series of components and optical and microwave amplifiers before and after the downconversion. 

The demodulated voltage $V_\mathrm{det}(t)$ represents the beating between the optical pump and optomechanical sideband. Being a beat note product, this signal is proportional to the amplitude of the pump and sideband electric field at the photo-detector input, i.e. $|V_\mathrm{det}|\propto |E_\mathrm{SB}|$. The power in the optomechanical sideband $P_\mathrm{SB}$ for a blue detuned pump is in turn proportional to $\gamma_\mathrm{om}n_\mathrm{m}$ where $n_\mathrm{m}$ is the average number of phonons in the mechanical mode which is assumed to be $\gg1$. We thus have that
\begin{equation}
    |V_\mathrm{det}|\propto |E_\mathrm{SB}|\propto \sqrt{P_\mathrm{SB}}\propto\sqrt{n_\mathrm{m}}.
\end{equation}

The resonant microwave tone can be viewed as directly driving the mechanical mode through the electromechanical coupling $\gamma_\mathrm{m,e}$. For a coherently driven mechanical mode we expect
\begin{equation}
    \label{eq:phonon_dynamics}
    n_\mathrm{m}(t) = n_\mathrm{i}e^{-\gamma_\mathrm{m}t} + n_\mathrm{f}(1 - e^{-\gamma_\mathrm{m}t/2})^2, \qquad \forall t>0.
\end{equation}
Here $n_\mathrm{i}$ is the initial phonon occupation prior to driving and $n_\mathrm{f}$ is the final phonon occupation after a long drive which is set by drive power, $\gamma_\mathrm{m,e}$ and total $\gamma_\mathrm{m}$. Note that the phonon ring-up follows different dynamics when driving coherently versus with a Markovian thermal bath. A thermal bath drives the mechanics with white noise leading instead to $n_\mathrm{m}(t) \propto 1-e^{-\gamma_\mathrm{m}t}$. This is commonly observed in OMC's at cryogenic temperatures where optical absorption induced noise acts as mechanical drive \cite{sonar_highefficiency_2025,kolvik_optomechanical_2025}.

With the above as background, we estimate $\gamma_\mathrm{m}$ from the bit array data by accessing the detected voltage amplitude through the quadrature sum $|V_\mathrm{det}| = \sqrt{I^2+Q^2}$ and fitting the result to the square-root of \eqref{eq:phonon_dynamics}:
\begin{align}
    \label{eq:fit_ringup}
    &|V_\mathrm{det}(t)| = V_\mathrm{f}(1 - e^{-\gamma_\mathrm{m}t/2}),\\
    &|V_\mathrm{det}(t)| = V_\mathrm{i}e^{-\gamma_\mathrm{m}t/2},\label{eq:fit_ringdown}
\end{align}
where we use \eqref{eq:fit_ringup} for ring-up and \eqref{eq:fit_ringdown} for ring-down with fitting parameters $V_\mathrm{i}$, $V_\mathrm{f}$.

\section{Qubit-to-mechanics swap}
\label{A7}
In the main text we consider a transducer where the mechanical mode is coupled to a microwave cable. If it instead were to be coupled to a superconducting microwave resonator, the resulting electromechanical coupling rate $\gem$ becomes \cite{jiang_lithium_2019}:
\begin{equation}
\begin{aligned}
    \gem&=\frac{1}{2}\sqrt{\gammame\om} \sqrt{\frac{Z_{\rm q}}{Z_0}} \quad \text{with} \quad Z_q=\frac{1}{\omega_\mu(C_{\rm IDT}+C_{\rm q})}. %\\
    %&=\frac{1}{2}\sqrt{\frac{\gammame}{Z_0(C_{\rm IDT}+C_{\rm q})}} 
\end{aligned}
\end{equation}
Here, $Z_{\rm q}$ is the impedance of the microwave resonator and $Z_{\rm 0}=50\ \Omega$ is the characteristic impedance of the microwave feedline. The microwave resonators frequency is $\omega_\mu$ and its capacitance is $C_{\rm q}$. The other quantities are explained in the main text. The coupling rate is boosted by the resonator's impedance. In the following we consider a transmon qubit, i.e. an anharmonic resonator. For a transmon qubit with $C_{\rm q}=70\ \rm fF$ in resonance with the mechanical mode, $Z_q=525\ \Omega$. Together with our measured (simulated) mechanical-to-microwave coupling, $\gammame$, we receive $\gemopi= 0.8\ \rm MHz$ ($3.6\ \rm MHz$).

To swap a photon from one to the other, the qubit would be tuned into resonance with the mechanical mode. For a qubit initialized in the excited state with the mechanical mode in the ground state Rabi oscillations occur at a frequency $2\gem$. To complete a full swap the swap rate should surpass the mechanical loss rate, giving the approximate condition $\gammami<4g$. For the measured (simulated) coupling rates above, this requires mechanical loss rates of $\gammami/(2\pi) < 3\ \rm MHz$ ($<14\ \rm MHz$ ). Given those values the electromechanical cooperativity, $\Cem=\frac{4\gem^2}{\gammam\kappamu}$, amounts to $\Cem=0.7$ ($\Cem=3$), where we assumed $\kappamu/2\pi=1.2\ \rm MHz$ \cite{meesala_nonclassical_2024}.
These limits should be considered an approximate lower bound at which swaps become possible. In practice, lower loss rates are desirable to also conduct the optomechanical up-conversion and achieve overall high fidelity for the protocol under consideration.

\end{document}